\documentclass[conference]{IEEEtran}
\IEEEoverridecommandlockouts
\usepackage{cite}
\usepackage{amsmath,amssymb,amsfonts}
\usepackage{algorithmic}
\usepackage{graphicx}
\usepackage{textcomp}
\usepackage{xcolor}
\def\BibTeX{{\rm B\kern-.05em{\sc i\kern-.025em b}\kern-.08em
    T\kern-.1667em\lower.7ex\hbox{E}\kern-.125emX}}
\begin{document}

\title{The Technologies Required for Fusing HPC and Real-Time Data to Support Urgent Computing}

\author{\IEEEauthorblockN{Gordon Gibb, Rupert Nash, Nick Brown, Bianca Prodan}
\IEEEauthorblockA{\textit{EPCC} \\
\textit{The University of Edinburgh}\\
Bayes Centre, 47 Potterrow, Edinburgh, EH8 9BT, UK \\
g.gibb@epcc.ed.ac.uk}
}

\maketitle

\begin{abstract}
The use of High Performance Computing (HPC) to compliment urgent decision making in the event of disasters is an important future potential use of supercomputers. However, the usage modes involved are rather different from how HPC has been used traditionally. As such, there are many obstacles that need to be overcome, not least the unbounded wait times in the batch system queues, to make the use of HPC in disaster response practical. In this paper, we present how the VESTEC project plans to overcome these issues and develop a working prototype of an urgent computing control system. We describe the requirements for such a system and analyse the different technologies available that can be leveraged to successfully build such a system. We finally explore the design of the VESTEC system and discuss ongoing challenges that need to be addressed to realise a production level system.
\end{abstract}


\section{Introduction}
The Visual Exploration and Sampling Toolkit for Extreme Computing (VESTEC) project is an ambitious EU-funded Horizon 2020 project that aims to fuse HPC with real-time data to support urgent decision making \cite{VESTEC}. VESTEC aims to build a flexible toolchain that incorporates the complete workflow required for an urgent computing event, from the data acquisition, through running simulations on HPC machines to presenting the results to an urgent decision maker.
This toolchain will be completely automated, so when a disaster begins it will be triggered (either by an urgent decision maker or sensors) to collect and process relevant data which is then used as input for simulations. The VESTEC system will then run a number of jobs on HPC machines to simulate the disaster. This could be a single simulation, a chain of simulations, or even an ensemble of simulations from which statistical inference can be made. The results of these simulations are then made available to the urgent decision maker so that they can use these to benefit the disaster relief. The system will be able to kick off new simulations or modify currently running ones according to new data from the ongoing disaster, or upon instructions from the urgent decision maker.

To build and evaluate the tools for such sophisticated workflows for urgent decision making, VESTEC is focused on three use cases. The first use case is forest fire modelling, which will incorporate data from satellite observations and up to date weather simulations to drive an ensemble of simulations that simulate the fire's progression, whose results are passed to the urgent decision maker. As new data arrives, new ensembles will be run to reflect the new data, and the urgent decision maker is also able to run new ensembles (for example to see how taking some action will affect the progression of the fire). This allows the urgent decision maker to select better evacuation strategies and fire suppression priorities. The second use case is that of simulating the spread of mosquito-borne diseases, which will incorporate data from weather forecasts, as well as land use data from satellites to run simulations predicting mosquito population's evolution with time, and hence proxies for the disease risks. The models can be driven and modified by the urgent decision maker to explore how strategies such as vector control alters the spread of the mosquitoes. This allows the urgent decision maker to choose appropriate vector control methods and the best allocation of medical supplies to those who are likely to need them the most. The final use case is that of space weather prediction. This will use data from observations of the Sun and from in-situ satellite measurements in the Earth's magnetosphere to simulate the changing magnetospheric field and the dynamics of high energy electrons in the magnetosphere. The urgent decision maker can use the results from these simulations to identify whether (and when) satellites should be powered down to protect their electronics, and if any actions need to be taken on the ground to protect electrical grids from large eddy currents induced by the earth's changing magnetic field that can damage transformers. 

In this paper, we will focus on the basic requirements for and possible technologies that can be used to construct the control system for an urgent computing framework such as is proposed by VESTEC. First, we will review the background in urgent decision making in Section \ref{background}. We then take an in-depth look at the required functionalities and components for such a system, and for each consider technologies that can be used as well as some design strategies in Section \ref{control_system}. In Section \ref{vestec_system} we describe the design decisions we have made for the VESTEC system, based on the analysis presented in Section \ref{control_system}. Finally, in Section \ref{discussion} we discuss the ongoing challenges that must be overcome to achieve a fully operating VESTEC system.

\section{Background} \label{background}
Disasters such as forest fires, disease outbreaks and extreme weather tend to evolve rapidly and require a rapid response to avoid, or at least mitigate loss of life and damage to property. It is therefore prudent that an urgent decision maker, the person or organisation responsible for managing the disaster relief, works with accurate information as quickly as possible for them to choose the best course of action. For example, in the case of a forest fire, if the urgent decision maker has up to date information on the location of the fire and the wind direction they can prioritise who to evacuate, as well as where to place firefighting teams to best combat the spread of the fire.

Advances in technology over the preceding decades have opened up many new opportunities in aiding urgent decision makers. In an ever more connected world, there are numerous data sources available, from satellites, IoT sensors, and even social media sources such as Twitter. There is a tremendous wealth of information in all this data if it can be mined and aggregate to aid the decision maker. Additionally, with the ubiquity of handheld, internet-connected computing devices such as mobile phones and tablets, it is possible for the decision maker to communicate instantly with rescue workers at the site of the disaster and disseminate any appropriate data. The urgent decision maker themselves can also be at scene at the disaster and still have access to all the up to date data they require to make quick decisions. 

One such example of the use of recent technological advances is the assessment of destruction after hurricanes \cite{LIDAR}. MIT Lincoln Laboratory's Humanitarian Assistance and Disaster Relief Systems Group plan to use LIDAR to generate a map of Puerto Rico's topology. After a hurricane, another LIDAR survey is be carried out, and compared to the original baseline map. Algorithms are then used to determine the locations of damage, which allows disaster relief crews to quickly identify where they need to go to provide assistance. This use of technology represents a step change in capability because, presently this is achieved by driving cars about or flying a small aircraft over the island and taking photographs, then manually having to identify locations of damage. The inclusion of LIDAR allows data to be acquired much more quickly, hence focusing relief effort to where it is needed in a more timely manner.

Another example is the Next-generation Incident Command System (NICS) \cite{ncis,ncis2}, also developed by MIT Lincoln Laboratory in partnership with the California Department of Forestry and Fire. This is a web-based platform that allows disaster workers in the field to coordinate large-scale emergency responses. It consists of an incident map with annotations and an interactive whiteboard that contains data inputted by relief workers and external data sources. This provides the capability for individuals and groups to coordinate a response, collaborating on decision making and sharing information on the rapidly evolving situation.

With the rapidly increasing power of HPC machines, it is now possible to model the evolution of disasters faster than real-time. This is a very exciting prospect because it permits the disaster to be forecasted, giving the urgent decision maker foresight in how the disaster may unfold, allowing them to explore a range of possible response options and take more effective preemptive actions to mitigate the disaster. Often such forecasts must be fed with observational data which needs to be collected from various sources, pre-processed and then input into a simulation, typically running on an HPC machine. The results from this simulation must then be post-processed and delivered to the urgent decision maker so they can take the appropriate actions. The developing field of using computing to aid urgent decision makers is known as urgent computing\cite{urgentcomputing}. 

Perhaps the most well-known and ubiquitous application of urgent computing is that of weather forecasting. The vast majority of weather forecasting centres have dedicated HPC resources to run forecasts upon, and often have access to backup facilities \cite{MetOffice,SA} if their primary machine fails or if they need more computing resources during a particularly extreme weather event. For example, if a severe storm is likely to hit a region of a country, the forecasts will predict this storm hours, possibly days in advance. The centre may then decide to run further high resolution forecasts to better constrain the path of the storm, and the severity of the conditions during the storm. Using this data they can provide weather warnings to members of the public and businesses, and advise them on the best course of action to take. As the storm progresses, the forecasts and hence warnings can be updated in almost real-time to provide the best possible advice to people in the affected region, potentially saving lives, and mitigating the damage to the local economy. 

The above weather forecasting example is, unfortunately, a very idealised example of urgent computing. As stated, most weather forecasting centres have dedicated HPC machines to run these forecasts on. Furthermore, these centres are continually running forecasts that they know will complete in a bounded time. As such their supercomputers are constantly in use and streamlined pipelines are in place to collect and process observational data, run the forecasts then obtain the results. This is not the case for most disasters. Forest fires, for example, are isolated events that occur sporadically, whilst earthquakes or disease outbreaks tend to occur at random. It is therefore not economically viable to have a dedicated HPC machine constantly ready and waiting to run forecasts for these disasters, because such a machine would likely be idle for the vast majority of the time. Additionally, a single disaster response HPC machine could also cause problems if multiple disasters occurred simultaneously, as there would be no capacity to scale to larger computing resources if required. For most disasters, it is therefore prudent to use time on existing international, national or regional HPC systems. Unfortunately, HPC systems tend to be optimised for job throughput rather than individual job latency and as such there could be very long (hours, days possibly even weeks) wait times in the queue. For disasters unfolding very quickly over minutes or hours, this makes traditionally configured HPC machines essentially useless as the results for a simulation may only be available long after the disaster has finished. 

The Special PRiority and urgent computing Environment (SPRUCE), is one approach to urgent computing that tries to overcome the limitations of batch systems for urgent computing \cite{SPRUCE}. With SPRUCE, urgent decision makers were given special tokens that could be used to raise a job's priority in a batch queue on an HPC system. Tokens came with different priorities, such as simply increasing the priority of a job in the batch queue, putting a job to the front of the queue, or suspending currently running jobs to immediately run the requested job. The priority chosen by the user depended on the urgency of the situation and the higher priority the more costly the use of the token became. This approach overcame the problem of long batch queue wait times, however it required an agreement with the HPC facility to enable such a policy, and modifications to the batch system to facilitate the use of SPRUCE tokens. Furthermore, there needed to be human intervention to use a token (and indeed to choose the level of queue priority required), as well as human intervention once a simulation has been completed to pull back any data required to produce a forecast for the urgent decision maker. 

\section{Technologies Required for an Urgent Decision Making Control System} \label{control_system}
\begin{figure}
    \centering
    \includegraphics[scale=0.8]{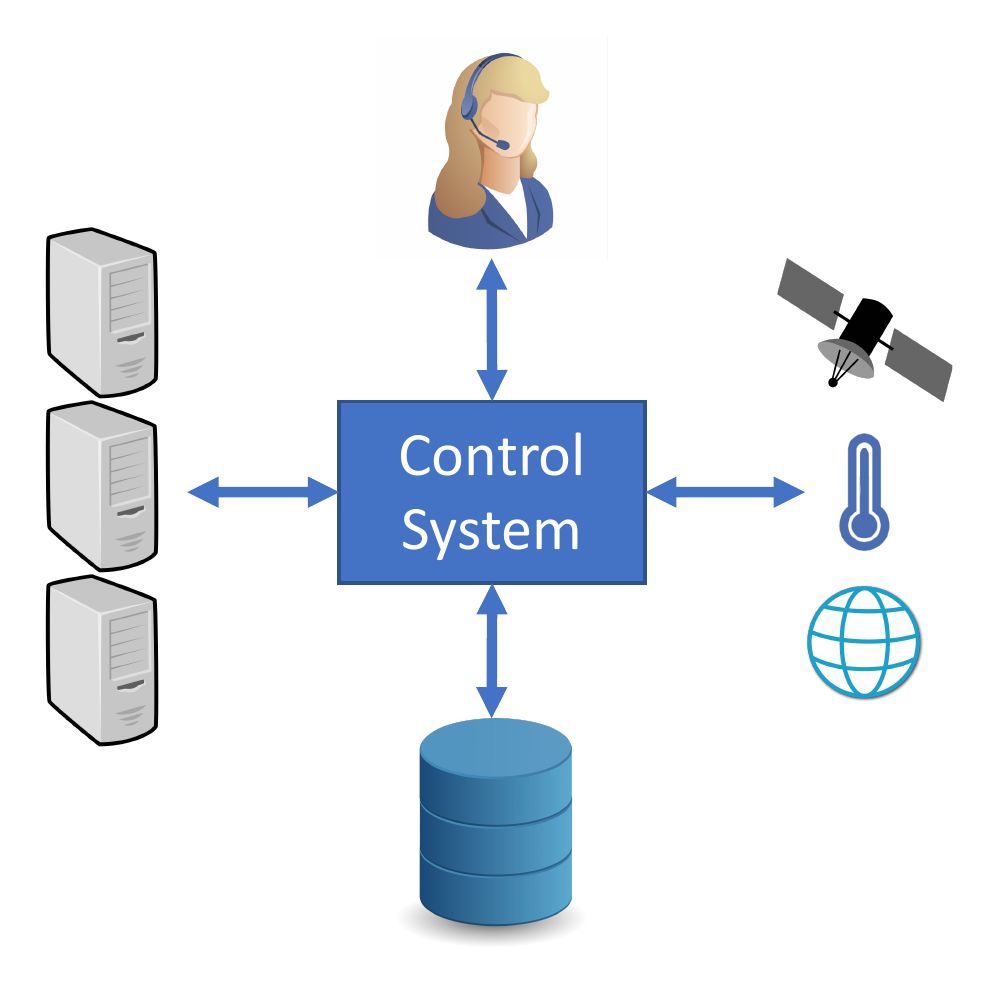}
    \caption{A schematic of a control system and how it should interact with the urgent decision maker (top), obtain data from data sources (right), read and write data to database(s) and other persistence stores (bottom) and connect to HPC machines (left).}
    \label{fig1}
\end{figure}

A centralised control system is the functionality that links the HPC machines, data sources and urgent decision makers together. Such a system must be capable of obtaining data from external data sources such as sensors, run jobs on an HPC machine to carry out the forecasts, and relay the results of the simulations to the urgent decision maker (Figure \ref{fig1}). It is vital that the control system tries to run jobs as quickly as possible to minimise the wait time between a job being submitted and the results being available. Ideally this system will have access to many HPC machines such that:
\begin{itemize}
    \item Jobs are able to run more quickly by picking the appropriate machine, hitting the optimal sweet spot of suitability for the target application, low queue time, and limited data movement
    \item The impact of an urgent decision workload can be evaluated and tracked on a machine by machine basis. As responding to disasters could involve running many jobs over many hours, it might be prudent to spread the workload across many machines, but such an approach should not significantly impact the operation of individual machines.
    \item The health of the individual HPC machines can be monitored and their availability tracked. If an individual machine becomes unavailable then the system can avoid submitting jobs to it in future, and previously scheduled jobs that were waiting to run can be allocated elsewhere.
\end{itemize}

We also identified that the control system should be written in a language that is powerful enough to express the complex abstractions that will inevitably be required. Ideally the language would also have access to numerous software libraries so that existing technologies can be leveraged easily. We identified that languages with web-based enterprise level functionality, such as Java and Python, are good candidate choices. Whilst the use of Java or Python might seem like a strange choice in the context of HPC, much of the control system itself is marshalling and control. Adapting such a language allows one to take advantage of its rich ecosystem and also enables the rapid prototyping and modification of functionality as lessons are learnt during development. Crucially, both languages have support for calling functionality in other languages, such as C or C++ \cite{javac,pythonc}. As such, if during development it is found that aspects of the control system are performing poorly and acting as a bottleneck in general, then these can be converted into a natively compiled language and still integrated with the overall control system. By first developing these components in a higher-level language the main structure and functionality of these components will have already been discovered, so any conversion at the language level should be a fairly trivial task.

In this paper we will now explore the major components that such a control system requires, and the technologies that could be leveraged in implementing these. Firstly, it is important to define some terminology we will be using. An \emph{activity} is a complete workflow for dealing with a specific disaster, whilst a \emph{job} is a particular part of that workflow running on an HPC machine. Over the course of an activity, many jobs may need to be run, in serial, concurrently, or a combination of the two depending on the workflow. For example, input data may need to be pre-processed before being used as input to a simulation, and both the pre-processing and simulation could be separate jobs, possibly on different HPC machines. 

The control system will require some persistent data store to contain simulation data and the current state of the System. It must also contain the functionality to connect to HPC machines to submit jobs, transfer data and monitor their queues. Data from external sources must be available to the control system, and it must be able to execute and organise the complex workflows involved in running activities. Finally, it must have a user interface such that the urgent decision maker can view results and steer a simulation if necessary.

\subsection{Persistence} \label{persistence}
A control system is required to store information on its state, namely the state of activities and their simulations, but also potentially input data and results from simulations. Here we will consider two possible technologies, namely relational databases and object storage, which can meet these requirements. 

Relational databases are ordered into one or more tables of rows and columns. Each row has a unique identifier key and a row can be linked to rows in other tables through a column that contains the key for the row in another table. This allows one to construct complex relations between tables. For example, there could be a table for activities, a table for individual jobs, and a table for the HPC machines. Each row of the job table would contain the key for an activity row, thus linking jobs to activities, as well as a key to a machine row, linking the job to a particular machine. One could then easily find all jobs associated with a particular activity running on a particular machine by searching for that given activity's key in the job table, and then refining this search by jobs whose machine key corresponds to the requested machine. There numerous implementations of relational databases, for example SQLite\cite{sqlite}, MySQL\cite{mysql} and MariaDB\cite{mariadb}. Most relational database implementations use the Structured Query Language (SQL) to retrieve data from the databases. Although the underlying implementation may be different, writing queries to obtain information is thus largely agnostic of the underlying implementation.

One could further abstract retrieving data from the database using an Object Relational Mapper (ORM)\cite{orm} which abstracts interaction with the database to a set of persistence objects, such that developers can work with the content of a database in the form of objects rather than raw SQL queries.
This permits one to integrate the database into the language being used for the control system, without having to explicitly run SQL queries to set or retrieve data.
To persist items into the database, the programmer sets the value of fields in an object and any links between objects to form relationships. This is exactly how developers tend to use normal objects in Python, the only difference being that the underlying ORM technology then automatically stores this information and any object relationships in the database. To retrieve data from the database the programmer will write queries, but unlike SQL queries these tend to be within the context of the programming language. The data is then returned as a populated set of objects that the code can work with, any modifications made to field values then automatically reflected in the database.

Relational databases are best suited for storing structured data, however are not so well suited for storing large amounts of unstructured data, for example simulation output files. For the large (many GBs) of simulation output data, an object store is a more appropriate technology. Object storage stores data as objects identified by a unique identifier which is used to retrieve the data. This is opposed to filesystem storage, which stores data in a file hierarchy. Each object in the object store tends to have a HTTP URL associated with it, so this data can be accessed over the HTTP protocol. Data used as input to simulations, or as output from simulations can be stored in object storage, with its unique identifier/URL stored in the relational database. This data can thus be retrieved as required, but not take up significant space in the database. The object store could be local to an HPC machine, part of the control system, or even in the cloud. It may be possible to have several object stores in different locations to benefit different requirements for different activities.

\subsection{HPC Machine Interface} \label{HPCinterface}
There must be a means for the control system to connect to HPC machines. Connectivity broadly falls under two main categories: submitting jobs and retrieving their results, and monitoring the machine status and queue. In this section we first describe the basic requirements to connect to an HPC machine, then discuss the specific requirements for each category. 

Connection to most HPC machines is achieved through the Secure Shell (SSH) protocol, whilst transferring data to or from  the machines is achieved through SSH file transfer protocols (SFTP) or secure copy (SCP) protocols. A software layer is thus required that is able to connect to a machine using SSH and SFTP/SCP. There exist a number of libraries that permit such functionality, for example with Python one can use SAGA\cite{saga}, Paramiko\cite{paramiko} or Fabric\cite{fabric}. The control system must have login accounts and passwords/SSH keys to gain access to the HPC machines and many HPC machines provide explicit access policies which often prohibit activities such as the sharing of accounts. A control system requires credentials for the appropriate machines, most likely in the form of its own machine processing user. This could cause issues from a policy perspective because connecting to an HPC machine and running jobs under a single user from the control system across many urgent decision makers might fall foul of account sharing policies. This important issue must be kept in mind when developing such a system, and can be mitigated through negotiation with the affected HPC centres to relax their policy/create an exception, or to have multiple machine accounts, one per decision maker. 

\subsubsection{Job Submission}
The vast majority, if not all, HPC machines use batch systems to submit jobs. There are a number of batch systems commonly used by HPC systems such as PBS, Slurm, Torque etc. As the control system will need to interact with several different HPC resources, each with potentially a different batch system, the control system must be aware of the different batch systems and how to use each one. Whilst the different batch systems differ in their command syntax, the general inputs to a batch command are similar between batch systems. For example, all require the node/core count, walltime and an accounting code, although the format of these may be different. It should therefore be possible to write a generic wrapper for submitting a job, taking these aforementioned variables as inputs, which then parses these into a submit command specific to the HPC machine. It should be mentioned that libraries such as SAGA and Fabric do possess limited functionality for batch job submission, although we have found that the expressibility they provide is not sufficient for a control system, which is not surprising given the novel use of HPC this represents.

In addition to submitting jobs, the HPC interface must also be capable of transferring input files and data onto the HPC. Similar to above, functionality already exists in SAGA, Paramiko and Fabric to provide SCP and SFTP file transfer, although it may be worth writing an interface layer to abstract lower level communication functions to higher level data transfer commands. If data is on a remote object store, it would be more efficient to transfer the data straight from the object store to the HPC machine, rather than onto the control system then onto the HPC machine. A file transfer abstraction layer could simplify file transfers for the developer by making this agnostic of the source and destination of files to be transferred. 

\subsubsection{Machine Status}
The HPC interface must also be able to query a machine's queue for the status of jobs running on it. This includes jobs submitted by the control system, but also the status of all jobs so that the control system can monitor the busyness of the queue and the status of the HPC machine more generally. As discussed above, HPC machines may use different batch systems so the commands that must be used to query the queue and their output will differ, however fundamentally the same information must be provided and returned. It should thus be possible to write an abstraction layer to query the queue status from a generic HPC machine, which at a lower level uses the appropriate commands and parses their output appropriately to obtain the queue information.

Querying the statuses of jobs submitted by the control system would be reasonably straightforward. The IDs of the submitted jobs would be passed to the batch system and it would return their statuses (e.g. Queued, Running, Complete, Error, etc). Based on this information the control system can then decide on whether further action is necessary, for example, if a job had completed the control system can then plan to transfer results off the HPC machine or submit a new job as required. 

A much more difficult yet interesting problem is that of querying the status of the entire queue. The status and properties of \emph{all} the jobs on the HPC machine in \emph{all} queues on that machine must be returned. This information must be collected for \emph{all} machines. The control system can then track all the jobs currently in a queue, queued and running, to gain an understanding of how busy the machine is and construct an estimated wait time for a possible job to be submitted. Considering all queues on all machines, the control system can then select the machine with the shortest queue time to submit a job to.

Whilst considering the current state of a queue is useful, it is far more powerful to collect historical information on each machine and their queues. The control system can periodically, for instance every 10 minutes, collect queue information and store this in its database. This allows the control system to model queue behaviours, and hence refine its predictions for estimated wait times. One challenge is that, for jobs currently running, the na\"{i}ve way to determine their remaining runtime is to compare how long they have currently been running for to their requested walltime. We found that when submitting jobs users tend to overestimate the required walltime, so using the requested walltime is always an overestimate of the remaining runtime in the queue. With access to historical job data, the control system can use this to statistically determine a more likely remaining runtime for jobs, and hence improve its predictions.

Storing historical data also allows the control system to highlight anomalies, for instance, if the machine suddenly becomes busy every day at a specific point in time, or whether the current machine load is abnormal. Additionally, this historical data enables a reliability metric to be associated with each machine, based on how often they are offline, which can be borne in mind when submitting jobs. Whilst this is a fairly crude approach to tracking machine health, it provides some level of understanding whether the job is likely to run through to completion or not and in the future can be extended to capture whether specific jobs have failed.

It would be possible to use a machine learning approach to consider the data for each queue on each machine and use this to predict wait times etc (e.g. see \cite{HPCML1,HPCML2}). It may also be possible to use a reinforcement learning\cite{reinforcementlearning} approach to improve the prediction power of the system as new queue data becomes available. Producing such a prediction system is likely a reasonable amount of work, however, such a system is a benefit to the wider HPC community, not just for urgent decision making, so doing such work would be very worthwhile.

\subsection{Sensor Interface}
The control system must be able to access data relevant to the disaster. This can come in many forms, from satellite data to weather forecasts to social media posts. Irrespective of its origin, the data needs to be used both as an input to simulations, but also potentially as a trigger to start an activity, or to modify an existing activity, for example if circumstances have changed drastically. Although we call this component the sensor interface, it is important to note that it need not include just sensors, but generally any interface to data external to the control system and HPC machines. Similarly, a \emph{sensor} in this case is a provider of data, and may not represent an actual sensor.

There are two approaches that can be used to obtain data, namely push-based and pull-based approaches. In the push-based approach, an external source notifies the control system of new data or of any significant change in the data. Technologically, this would require the control system to export a REST interface\cite{rest} that the external sensor can send HTTP requests to with the new data. The sensor must also be aware of the control system and know when it needs to push data to it. In the pull-based approach, the control system will periodically poll sensors for new data, likely using an HTTP GET request.

Upon obtaining data, the Sensor Interface must be able to trigger events or actions for the control system to carry out. For example this could be running processing work on the data to reduce it to a useful form for simulations or to initiate off new jobs or activities. Simple processing could be carried out on the control system, but in principle this processing could be deferred to an HPC machine as a job.

\subsection{Simulation Manager} \label{simmanager}
The simulation manager can be considered the heart of any such control system. It is responsible for determining which simulations must be run, where they need to be run, and in which order to complete an activity. All the components mentioned previously are thus tools that are used by the simulation manager to obtain data, set up and run jobs on HPC machines. The simulation manager must be aware of the various activity types (e.g. forest fire, disease outbreak or space weather event) and how to progress these activities. Each activity could, in principle, have a quite complex workflow, and so it is vitally important that the simulation manager has the capability to understand and execute these workflows. 

There are a variety of workflow technologies available, and one must consider which one is most suitable for use in urgent computing. From surveying the technological landscape it was found that the two most mature and ubiquitous approaches are the Common Workflow Language (CWL) and Apache Taverna. A major benefit to both these technologies is that, not only are there existing implementations of workflow engines that can be leveraged, but also a wealth of other supporting tools to perform tasks such as composing workflows and visualising them. 

The Common Workflow Language (CWL) is not a specific implementation per-se, but instead an open standard for describing workflows \cite{cwl}. The goal is to encode this information in a way which makes the workflows portable and scalable across a variety of software and hardware environments. Driven by data-intensive science, and specifically the life sciences, CWL is developed by a working group comprising of members across the data workflow community. This standard is becoming dominant in the workflow community and workflow technologies that followed different description formats have either moved to CWL or at-least support CWL based descriptions.

By contrast, Apache Taverna\cite{taverna} is an actual software suite for designing and executing workflows. This is an Apache incubator project, meaning that it is currently not mature enough to be a full Apache project but is being supported by the foundation to move to that status. This technology has also been developed from life sciences and is currently in use with some of the large organisations in that area such as the National Center for Biotechnology Information, the European Bioinformatics Institute, and the DNA Databank of Japan. Oriented around a workbench, users can use their tools to design workflows on the desktop and explore them to the Taverna engine for execution either locally or remotely. 

Considering the possible workflows required to carry out an urgent computing activity, we concluded the following four criteria that a workflow system must possess:
\begin{enumerate}
    \item Support concurrent execution of different parts of the workflow at different levels in the hierarchy. 
    \item The ability to execute multiple, different, workflows concurrently and progress these independently from one another.
    \item Reasonable performance characteristics, whilst these are not computationally intensive it is important that the workflow engine is able to progress workflows in a timely manner.
    \item Support conditional branching in workflows, where different routes will be taken depending upon specific pre-conditions.
\end{enumerate}
CWL does not support the fourth criterion of conditional branching, which could form an important part of activity workflows. Apache Taverna does support conditional workflows to some extent, although this does not seem like a commonly used feature, but it is a heavy weight technology and from testing has raised doubts around criterion 3. The other danger of Taverna is that, due to its incubator status, it is currently in-flux, with the overarching workflow description language having changed recently and there are no guarantees that the Apache foundation will bring it to full project status. As such, Taverna represents a significant risk as it could change substantially or even cease to exist in its current form. Whilst the above analysis seems to exclude both of the considered workflow technologies as they do not meet our criteria, it may still be possible to leverage the workflow technologies to describe specific parts of the workflow, rather than the workflow as a whole. It is required, however, to write a custom workflow solution for the overarching workflows of different activities.

\subsection{User Interface}
There must be an interface to the control system and in particular, the simulation manager, so that an urgent decision maker can track the progress of simulations, modify their parameters, quickly view results and initiate new simulations and activities as required. Given the ubiquity of web applications today, it is reasonable to create a web-based interface to the control system, as a plethora of web technologies that can be leveraged exist. For more detailed visualisation of simulation results, a technology such as Paraview\cite{paraview} or VisIt\cite{visit} may be required, however a simple web interface suffices for the majority of requirements. The advantage of a web interface is that the only software requirements on the side of the urgent decision maker is a modern web browser. This permits the urgent decision maker to gain access to the control system whilst out in the field with hardware such as a mobile phone or tablet, thereby freeing them from requiring to be close to a reasonably powerful workstation. 

An HTML document is static (e.g. unchanging), however this can be made interactive through the use of JavaScript which permits the web page to be dynamically updated. Any such web interface will have to provide information which can vary based on the website user, any ongoing activities, and various other factors. This variability in information can be achieved either by the web interface serving dynamically created HTML documents on the fly as the user requests them, or by the web interface serving a static HTML page with embedded JavaScript that then fetches and displays the required data. This data would be retrieved (on the server side) from the relational database.

As previously mentioned, a web interface is sufficient for basic tasks, however fails for visualisation of complex simulation output. For this, more complex specialist tools are required which are able to display remote data. One such tool is Paraview, which can connect to a remote HPC machine and be used to display simulation data. Although ideal for visualisation, these tools are not suitable for use as an interface to the control system. Access to the control system and Visualisation of data therefore require both a web interface and visualisation software such as Paraview. The Web interface could be used to provide a handle to the data which can then be displayed by the visualisation software, however. For example the web interface could provide the user with an IP address, port, and any credentials required to connect the visualisation software to a running simulation or to display remote results.

\section{The VESTEC Control System} \label{vestec_system}
\begin{figure}
    \centering
    \includegraphics[scale=0.85]{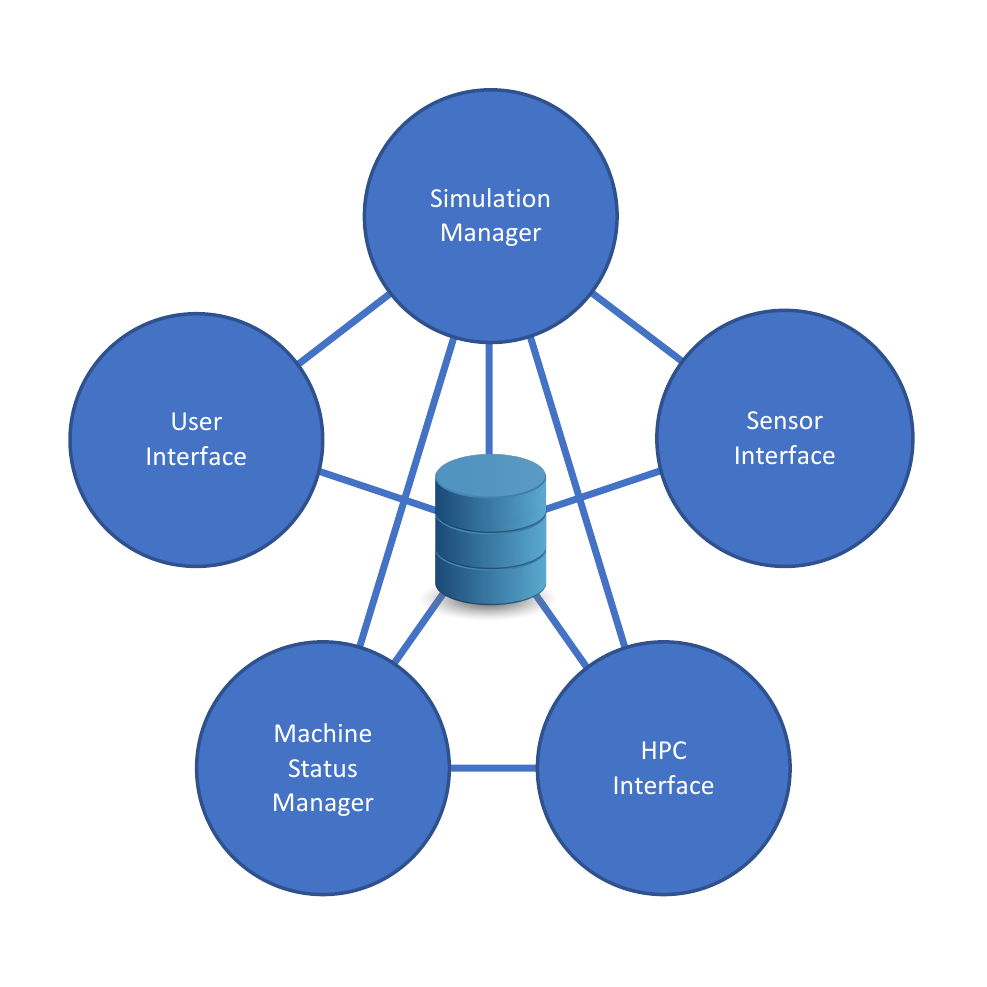}
    \caption{A schematic of the VESTEC control system. Note that every component is connected to the database (centre) and to the simulation manager (top), but not all components are necessarily connected to each other.}
    \label{fig2}
\end{figure}

In this section we will describe the design and technology decisions we have made as part of the VESTEC project given the discussion in Section \ref{control_system}. 
Firstly, we chose to develop the control system in Python, namely Python 3. Python was chosen because of its ease of use, large ecosystem of libraries and the ease at which it can be integrated with C/C++ if necessary. We use Python 3 because Python 2 planned for returement at the end of 2019\cite{python2}, although we note that this slightly reduces our choice of libraries to use as several have not been ported to Python 3. From Section \ref{control_system} it is clear that the control system naturally partitions into several separate components, such as the database and persistent storage, HPC machine interface, sensor interface, web interface and workflow manager. We therefore chose to create separate components which communicate to each other. Structuring the system like this allows each component to be tested in isolation, and also permits different people/teams to work on each component individually. For deployment, we use Docker\cite{docker,dockerrev} containerisation whereby each component resides within a Docker container, and the whole control system is run using \emph{docker compose}\cite{dockercompose}. Use of containerisation allows the system to be developed on a local machine but then easily ported to a server without having to worry about having a specific software stack or configuration on that server. Figure \ref{fig2} displays a schematic for the VESTEC control system's components and how they are connected.

The individual components of the control system require to communicate with each other to coordinate and achieve the full functionality required of the system. To this end, they need to be able to communicate with each other through well-defined interfaces. At present we are using a RESTful\cite{rest} interface between components, using the Python Flask web microframework\cite{flask}, whose simplicity and flexibility is of major benefit here. Each component exposes a RESTful interface, with messages being sent with HTTP requests, carrying any required information in JSON. Whilst REST has become a very popular choice it does have several disadvantages \cite{RestPerf} in the context of implementing the VESTEC system:
\begin{itemize}
    \item REST is blocking, such that it requires that the client receives a response from the service before it can proceed. Hence whilst naturally synchronous operations, for instance, waiting for resulting data from the server, are fairly simple, supporting asynchronous operations is not.
    \item REST is a text-oriented protocol, with data most often returned in JSON format. Over HTTP(S) this can become problematic due to the overhead of data transfer. 
    \item Clients must know exactly which services they need to connect to and access, for instance, it is not possible to broadcast a message out to many different web services and for an appropriate one to handle it. 
\end{itemize}
Therefore, in addition to a REST approach, within the VESTEC system a message oriented paradigm has been adopted. This is where messages are sent from a client to some broker asynchronously, and these are then forwarded onto the most appropriate handler. 

\begin{figure}
    \centering
    \includegraphics[scale=0.45]{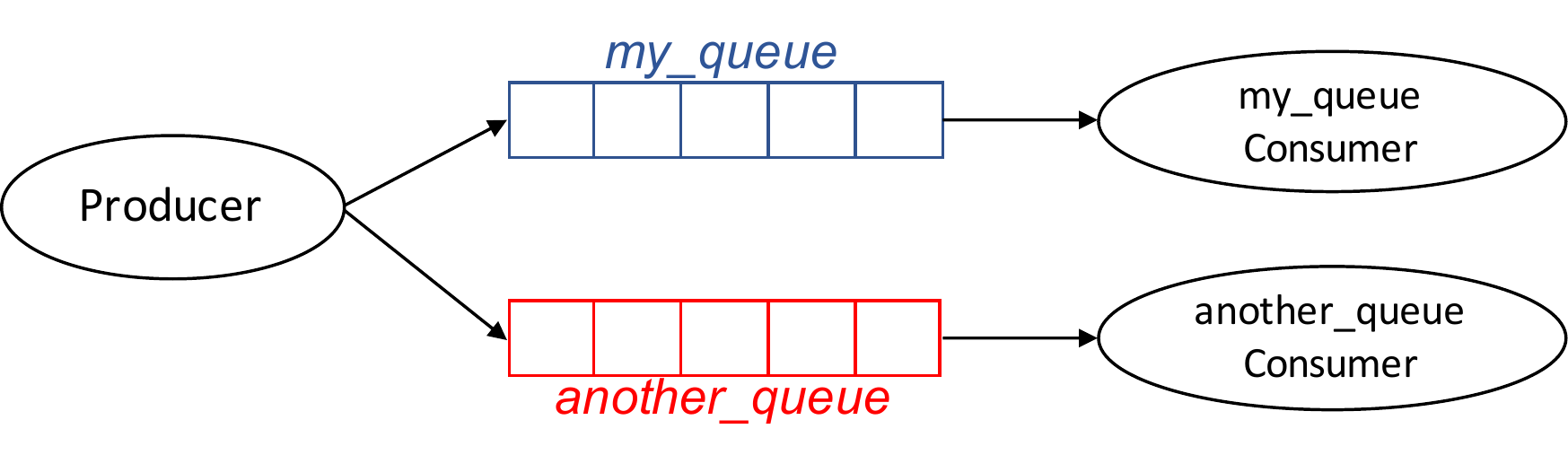}
    \caption{An illustration of the AMQP central concepts of producers, queues and consumers.}
    \label{amqp}
\end{figure}

Advanced Message Queueing Protocol (AMQP)\cite{amqp}, which is a very popular and standard way of driving interaction via messages, has been selected and various implementations of the protocol are available. In the general AMQP approach, a number of independent queues are declared, each with associated consumer(s) that will pick data from the queues in a specific order, often based on the first-in first-out method, and process this as appropriate. This is illustrated in Figure \ref{amqp}. Producers generate the messages and there can be any number of these interacting with the queues, sending data to different queues at different points if necessary. Messages in the queues are persisted so, for instance, if the system was to fail then data is not lost but instead will continue to be consumed once the system has been restarted. Such behaviour is important for a production version of the VESTEC system, where data loss in the event of failure could have severe consequences. 
AMQP supports multiple patterns of message routing and, as such, it is possible to define powerful interactions. For example, messages can be published to one or more queues, to which multiple consumers can subscribe, without the publisher having to be explicitly aware of this. One can also connect multiple consumers to the same queue which enables parallelism at the message consumption level, where multiple consumers execute concurrently consuming different messages in the same queue. Another major benefit of AMQP is that they can be easily scaled up across multiple servers to improve reliability or performance. The services also offer message buffering which means that they can cope with an influx of messages at a greater rate than the consumer(s) can process. 

One can also use AMQP much like a conventional HTTP request using a Remote Procedure Call (RPC) style of messaging where the original message is dispatched with a suitable reply-to metadata header and the consumer can then use this as the return address to answer the query. At the code level, for VESTEC the RabbitMQ\cite{rabbitmq} implementation of AMQP has been selected. Due to the simplicity of the technical solution, the VESTEC system is currently built on just RESTful web services (using Flask) internally. As the development phase progresses the plan is for a number of these to be replaced by AMQP. Whilst not all components of the VESTEC system will be connected via AMQP, because it is not necessary, this technology will play an important role for a number of components in the future.

For the database component of the VESTEC system, we are using a MariaDB database server. The different components of the system connect to the database in Python using the PonyORM\cite{pony} object relational mapper. As mentioned in Section \ref{persistence}, using an ORM abstracts the database technology away from the programmer. As such, using PonyORM means that we can change the database server from MariaDB to something else at a later date with minimal effort. For connecting to object storage, we use the Minio\cite{minio} Python client, which allows APIs to connect to any Amazon S3 API compatible object store. Given Amazon S3's ubiquity, using Minio permits us to be able to use a large number of object storage technologies at a later date.

For the HPC machine interface, we use Fabric and the lower level Paramiko library upon which Fabric is built. Fabric provides much of the functionality we need (as described in Section \ref{HPCinterface}) however not all. We, therefore, built abstractions upon this that implement the functionality we need. We also wrote a Python class that can interpret the output from PBS queues and store this in the database so that the queue can be monitored over large times. We have been collecting data on the queues of two of the machines we have access to, ARCHER\cite{archer} and Cirrus\cite{cirrus} for the previous four months and we plan to use this data to train a machine learning model to help predict the wait time for jobs. The HPC machine interface is structured as two components, the HPC machine interface, but also the machine status manager, which is responsible for analysing queue data to gain an insight on the queues to help with choosing which machine to run jobs on. We have taken the design decision to use AMQP to connect the HPC machine interface to the rest of the VESTEC system because the technology promotes an asynchronous way of working, which REST does not. This means that heavily used parts of the system, such as the HPC machine interface, can proceed asynchronously from other parts and will thus process and respond to requests in an orderly fashion, as marshalled by the active message queue, rather than acting as a bottleneck as other parts of the VESTEC system are blocking for a response. This is particularly important when, for example, we need to transfer a sizeable file to the HPC machine as we do not want the system to block until this is complete.

As discussed in Section \ref{simmanager} we found that existing workflow implementations are not flexible enough to meet the requirements of the simulation manager, however some could be leveraged to carry out parts of the workflow.
The approach chosen for the VESTEC control system is to use CWL, due to its standardised nature, but only use this to describe each individual activity of the workflow, and connect these using a bespoke implementation written for the VESTEC system. Whilst the intention was initially to avoid this, given the current state of these technologies it is the best option at this time. In fact, much of this can be abstracted through the use of AMQP, where queues are defined dynamically according to the different workflow rules and the production of a message occurs when a trigger fires, the corresponding rule looked up and the message is placed on a specific queue which is then consumed by the next action to execute. In this manner, the workflow engine builds upon AMQP, producing and consuming messages, along with determining how to route these which involves looking up which queue to place the message on. 

\begin{figure}
    \centering
    \includegraphics[scale=0.20]{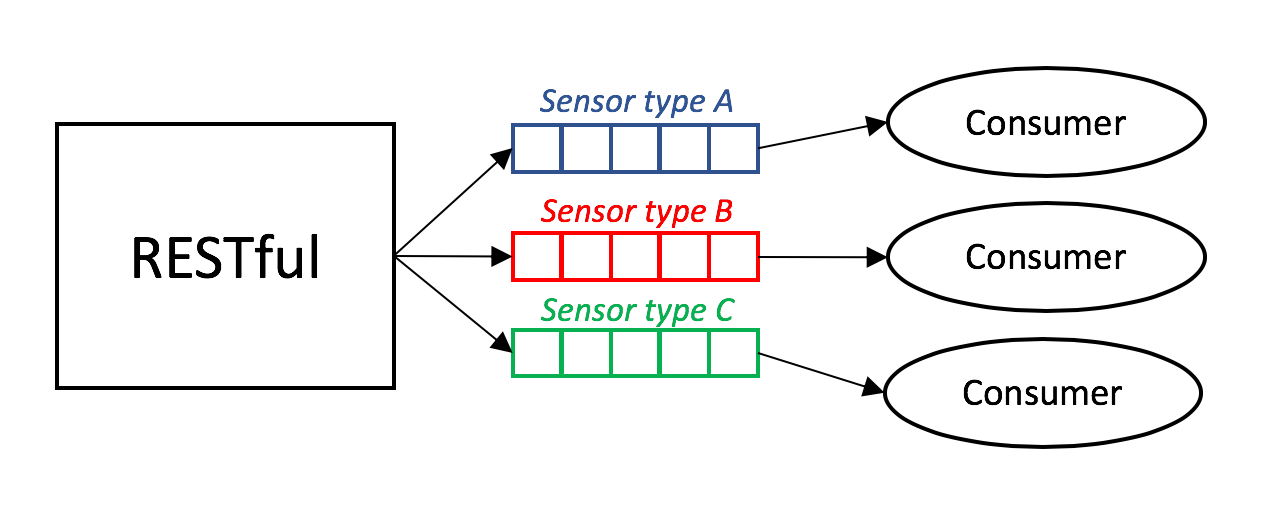}
    \caption{Illustration of the planned sensor approach where a RESTful approach is used to obtain data from sensors, then this is fed into appropriate AMQP queues for processing by consumers.}
    \label{sensors}
\end{figure}

The sensor interface is currently the least mature component of the VESTEC server. We plan to have a RESTful system (e.g. Flask) for any push-based requests from sensors, and another system to pull for data. Once these systems have acquired the data, they will act as an AMQP producer, publishing the received sensor data to the corresponding queue representing that type of sensor. Consumers, one for each sensor type, will listen to their queue and consume messages from them. This is illustrated in Figure \ref{sensors}. The role of these consumers will be to perform any sensor type-specific actions, such as data conversion. Whilst there is no plan to convert data to any common format, the meta-data will need to be standardised and it might be that only the meta-data is carried forward into the rest of the VESTEC system, with the consumer storing the actual data in an appropriate object store which can be retrieved at a later time as required. The consumer will then interact with the workflow manager to drive the handling of the arrival of this data. Orienting the majority of the complexity involved in processing sensor data within the AMQP approach means that consumers, which contain the majority of the work to process sensor data, are protected by the queues and-so if they fall behind, the messages simply queue up. AMQP consumers can also be configured to run concurrently, potentially across multiple machines, and the underlying AMQP implementation then handles this. Lastly, as described above, the AMQP implementation persists the messages in the queue, so if the VESTEC system fails for whatever reason, then sensor data is not lost but will start being delivered again once the system restarts.

We have adopted the approach of serving static HTML (with JavaScript) to the user's web browser for the web interface. The JavaScript then fetches data from the VESTEC system using HTTP requests via a defined API. Built on AJAX, this approach issues asynchronous requests to the VESTEC server by sending a request to a specific URI with optional input data and proving a callback function which is executed when a response is received from the server. We have structured the system with a NGINX server to serve the HTML, and a separate component running Flask to respond to the requests from the AJAX calls. With regards to accessing remote data directly for visualisation with Paraview, we have not implemented this yet, but we envisage the web interface presenting the user with an IP address, port number and any authentication necessary to connect their Paraview Client directly to the HPC machine. This prevents any remote connections from having to go through the VESTEC server, thus reducing the network load on the server and also removing the need for us to develop a means to enable forwarding such network traffic from the HPC machine to the user.

\section{Discussion and Conclusions} \label{discussion}
In this paper we have explored the requirements for a generic urgent computing control system. We have described the various components that this system must contain, as well as the possible technologies that could be used to construct the components and build the system as a whole. The system must consist of some form of permanent storage, incorporating a database to record its state and possible an object store for storing any sizeable data required by the system. It must also be able to connect to HPC machines, both for submission of jobs and retrieval of job results, but also to monitor queue status and machine health. A system must be able to retrieve data from data sources in order to prepare simulations and react to changes in circumstances of the disaster. It must be able to coordinate and execute the potentially complex workflows involved in collecting and pre-processing data, running simulations, post-processing simulation results, and possibly using these to run further simulations. Finally, it must provide an interface for the urgent decision maker to view results of simulations and to initiate/control simulations.

We then outlined the design decisions made in the VESTEC control system. We chose to write the system in Python, with the option to write some components/routines in C/C++ if we required the extra performance a compiled language provides. Each component is developed separately and communicates through RESTful interfaces, although we plan to use AMQP, namely RabbitMQ between some components to overcome some limitations of REST. The components run within Docker containers, and the full system is executed with \emph{docker compose} to allow easy development and deployment. For the database we have adopted use MariaDB, using the PonyORM object relational mapper to interface the Python code to the database. We use the MinIO Python API to connect the system to any object stores we will require. The HPC machine interface is based on Fabric and Paramiko to connect to HPC machines, submit jobs and transfer data. For workflow management/execution we plan to write our own system as no pre-existing workflow technologies meet our needs, however, the common workflow language may be used to express certain parts of our workflows where applicable. Finally, a HTML/JavaScript web page is used as the interface to the system, which can forward connection details to an HPC machine to the urgent decision maker if they require advanced results visualisation through Paraview. 

One challenge we have yet to overcome is that of how to steer currently running simulations. This is a desirable functionality as if a new unforeseen development occurs with the disaster, the simulations can be changed almost instantly to reflect this rather than having to submit new simulations, which may have to wait in a machine's queue for some time. This also could come in useful for the urgent decision maker to modify running forecasts to investigate the effects of a proposed course of action on the disaster's evolution, again avoiding the time penalty of having to wait in a queue if new jobs were to be submitted. In order to steer simulations, we would have the simulation codes to expose an interface from which we can communicate with it to inform it of changes. This could either be via some RESTful-like interface or via the HPC machine's filesystem. Given that some HPC machines' compute nodes are not connected to the internet, communication over filesystem seems to be a more likely approach. We will continue to research the best way to achieve this once work on the control system is nearly completed, however we note that much of this functionality will be required in the application codes, with the VESTEC control system only having to provide the urgent decision maker with a means to access the application code's interface.

Within the VESTEC project the control system will only federate over three HPC machines, ARCHER and Cirrus, both hosted by EPCC in Edinburgh, Scotland, and the Beskow\cite{beskow} system hosted by KTH in Stockholm, Sweden. Ideally, a system such as VESTEC could have access to many HPC machines to be able to balance the load of any large urgent computing requirements and to maximise the possibility of getting high throughput of urgent computing jobs. Unlike the approach adopted by SPRUCE, we do not use tokens to raise our queue priority, but rather federate over many HPC machines in the hope that some of these machines may have relatively short queue times. It remains an open question whether this is entirely sufficient, but a major benefit is that it doesn't require special arrangements with the HPC centres to gain queue-jumping abilities. That being said, it is definitely a possibility that the VESTEC system could be combined with a token-based system to allow us to run jobs more quickly. One alternative to using HPC machines is to use the cloud, however, this is not as attractive an option as it may initially seem. In a previous study for VESTEC\cite{vestecnick}, it was found that the performance of cloud systems tends to be significantly lower than a corresponding HPC machine, mainly due to filesystem performance, the cost was higher, and the cloud is not as elastic as one would hope, with reasonably long startup wait times for large, infrequent compute requirements. Whilst these make it impractical to rely upon a cloud-based resource, it would still be plausible to supplement the HPC machines we federate over with some cloud resources for some jobs/applications.

\section*{Acknowledgments}
This work was funded under the EU FET VESTEC H2020 project, grant agreement number 800904.


\bibliographystyle{./bibliography/IEEEtran}
\bibliography{./bibliography/IEEEabrv,./bibliography/vestec}

\begin{thebibliography}{10}
\providecommand{\url}[1]{#1}
\csname url@samestyle\endcsname
\providecommand{\newblock}{\relax}
\providecommand{\bibinfo}[2]{#2}
\providecommand{\BIBentrySTDinterwordspacing}{\spaceskip=0pt\relax}
\providecommand{\BIBentryALTinterwordstretchfactor}{4}
\providecommand{\BIBentryALTinterwordspacing}{\spaceskip=\fontdimen2\font plus
\BIBentryALTinterwordstretchfactor\fontdimen3\font minus
  \fontdimen4\font\relax}
\providecommand{\BIBforeignlanguage}[2]{{%
\expandafter\ifx\csname l@#1\endcsname\relax
\typeout{** WARNING: IEEEtran.bst: No hyphenation pattern has been}%
\typeout{** loaded for the language `#1'. Using the pattern for}%
\typeout{** the default language instead.}%
\else
\language=\csname l@#1\endcsname
\fi
#2}}
\providecommand{\BIBdecl}{\relax}
\BIBdecl

\bibitem{VESTEC}
\BIBentryALTinterwordspacing
(2018) {Visual Exploration and Sampling Toolkit for Extreme Computing}.
  [Online]. Available: \url{https://vestec-project.eu}
\BIBentrySTDinterwordspacing

\bibitem{LIDAR}
\BIBentryALTinterwordspacing
(2018) Using lidar to assess destruction in {P}uerto {R}ico. [Online].
  Available:
  \url{http://news.mit.edu/2018/mit-lincoln-laboratory-team-uses-lidar-assess-damage-puerto-rico-0830}
\BIBentrySTDinterwordspacing

\bibitem{ncis}
\BIBentryALTinterwordspacing
(2018) {L}incoln {L}aboratory honored for providing access to technology for
  disaster response. [Online]. Available:
  \url{https://www.ll.mit.edu/news/lincoln-laboratory-honored-providing-access-technology-disaster-response}
\BIBentrySTDinterwordspacing

\bibitem{ncis2}
\BIBentryALTinterwordspacing
(2014) {Next Generation Incident Command System Fact Sheet}. [Online].
  Available:
  \url{https://www.dhs.gov/publication/st-next-generation-incident-command-system-fact-sheet}
\BIBentrySTDinterwordspacing

\bibitem{urgentcomputing}
\BIBentryALTinterwordspacing
S.~H. Leong and D.~KranzlmÃŒller, ``Towards a general definition of urgent
  computing,'' \emph{Procedia Computer Science}, vol.~51, pp. 2337 -- 2346,
  2015, international Conference On Computational Science, ICCS 2015. [Online].
  Available:
  \url{http://www.sciencedirect.com/science/article/pii/S1877050915012107}
\BIBentrySTDinterwordspacing

\bibitem{MetOffice}
\BIBentryALTinterwordspacing
(2017) The {M}et {O}ffice supercomputer is one of the most powerful in the
  world dedicated to weather and climate. [Online]. Available:
  \url{https://www.metoffice.gov.uk/about-us/what/technology/supercomputer}
\BIBentrySTDinterwordspacing

\bibitem{SA}
\BIBentryALTinterwordspacing
(2017) {CSIR} and {SA} weather services partner for the development of weather
  and climate products and services. [Online]. Available:
  \url{https://www.csir.co.za/csir-and-sa-weather-services-partner-development-weather-and-climate-products-and-services}
\BIBentrySTDinterwordspacing

\bibitem{SPRUCE}
P.~Beckman, S.~Nadella, N.~Trebon, and I.~Beschastnikh, ``Spruce: A system for
  supporting urgent high-performance computing,'' in \emph{Grid-Based Problem
  Solving Environments}, P.~W. Gaffney and J.~C.~T. Pool, Eds.\hskip 1em plus
  0.5em minus 0.4em\relax Boston, MA: Springer US, 2007, pp. 295--311.

\bibitem{javac}
\BIBentryALTinterwordspacing
(2014) {How to call a C program from Java?} [Online]. Available:
  \url{https://javapapers.com/core-java/how-to-call-a-c-program-from-java/}
\BIBentrySTDinterwordspacing

\bibitem{pythonc}
{M. F. Sanner}, ``Python: a programming language for software integration and
  development,'' \emph{J Mol Graph Model}, no.~1, 1999.

\bibitem{sqlite}
\BIBentryALTinterwordspacing
{SQLite}. Accessed 2019-08-21. [Online]. Available:
  \url{https://www.sqlite.org}
\BIBentrySTDinterwordspacing

\bibitem{mysql}
\BIBentryALTinterwordspacing
{MySQL}. Accessed 2019-08-21. [Online]. Available: \url{https://www.mysql.com}
\BIBentrySTDinterwordspacing

\bibitem{mariadb}
\BIBentryALTinterwordspacing
{MariaDB.org - Supporting continuity and open collaboration}. Accessed
  2019-08-21. [Online]. Available: \url{https://mariadb.org}
\BIBentrySTDinterwordspacing

\bibitem{orm}
\BIBentryALTinterwordspacing
S.~Wambler. (2006) {Mapping Objects to Relational Databases: O/R Mapping In
  Detail}. [Online]. Available:
  \url{http://www.agiledata.org/essays/mappingObjects.html}
\BIBentrySTDinterwordspacing

\bibitem{saga}
G.~Iwai, Y.~Kawai, T.~Sasaki, and Y.~Watase, ``{SAGA-based user environment for
  distributed computing resources: A universal Grid solution over
  multi-middleware infrastructures},'' \emph{Procedia Computer Science}, no.~1,
  2010.

\bibitem{paramiko}
\BIBentryALTinterwordspacing
{Paramiko}. Accessed 2019-08-21. [Online]. Available:
  \url{http://www.paramiko.org}
\BIBentrySTDinterwordspacing

\bibitem{fabric}
{A. Hannah}, ``Fabric: a system administrator's best friend,'' \emph{Linux
  Journal}, no. 226, 2013.

\bibitem{HPCML1}
J.~Guo, A.~Nomura, R.~Barton, H.~Zhang, and S.~Matsuoka, ``Machine learning
  predictions for underestimation of job runtime on hpc system,'' in
  \emph{Supercomputing Frontiers}, R.~Yokota and W.~Wu, Eds.\hskip 1em plus
  0.5em minus 0.4em\relax Cham: Springer International Publishing, 2018, pp.
  179--198.

\bibitem{HPCML2}
{V. Jancauskas and T. Piontek, P. Kopta and B. Bosak }, ``Predicting queue wait
  time probabilities for multi-scale computing,'' \emph{Philosophical
  transactions. Series A, Mathematical, physical, and engineering sciences},
  no. 377, 2018.

\bibitem{reinforcementlearning}
\BIBentryALTinterwordspacing
L.~P. Kaelbling, M.~L. Littman, and A.~W. Moore, ``Reinforcement learning: {A}
  survey,'' \emph{CoRR}, vol. cs.AI/9605103, 1996. [Online]. Available:
  \url{http://arxiv.org/abs/cs.AI/9605103}
\BIBentrySTDinterwordspacing

\bibitem{rest}
L.~Richardson and S.~Ruby, \emph{RESTful web services}.\hskip 1em plus 0.5em
  minus 0.4em\relax O'Reilly Media, 2008.

\bibitem{cwl}
B.~Chapman, J.~Chilton, M.~Heuer, A.~Kartashov, D.~Leehr, H.~M{\'e}nager,
  M.~Nedeljkovich, M.~Scales, S.~Soiland-Reyes, and L.~Stojanovic,
  \emph{\BIBforeignlanguage{English}{Common Workflow Language, v1.0}},
  P.~Amstutz, M.~Crusoe, and N.~Tijanić, Eds.\hskip 1em plus 0.5em minus
  0.4em\relax United States: figshare, 7 2016, specification, product of the
  Common Workflow Language working group. http://www.commonwl.org/v1.0/.

\bibitem{taverna}
D.~Hull, K.~Wolstencroft, R.~Stevens, C.~Goble, M.~Pocock, P.~Li, and T.~Oinn,
  ``{Taverna: a tool for building and running workflows of services},''
  \emph{Nucleic acids research}, no.~34, 2006.

\bibitem{paraview}
\BIBentryALTinterwordspacing
{Welcome to Paraview}. Accessed 2019-08-21. [Online]. Available:
  \url{https://www.paraview.org}
\BIBentrySTDinterwordspacing

\bibitem{visit}
\BIBentryALTinterwordspacing
{VisIt}. Accessed 2019-08-21. [Online]. Available:
  \url{https://hpc.llnl.gov/software/visualization-software/visit}
\BIBentrySTDinterwordspacing

\bibitem{python2}
\BIBentryALTinterwordspacing
(2008) {PEP 373 -- Python 2.7 Release Schedule}. [Online]. Available:
  \url{https://www.python.org/dev/peps/pep-0373/}
\BIBentrySTDinterwordspacing

\bibitem{docker}
\BIBentryALTinterwordspacing
{Enterprise Container Platform for High-Velocity Innovation}. Accessed
  2019-08-21. [Online]. Available: \url{https://www.docker.com}
\BIBentrySTDinterwordspacing

\bibitem{dockerrev}
C.~Boettiger, ``{An introduction to Docker for reproducible research},''
  \emph{ACM SIGOPS Operating Systems Review}, no.~41, 2015.

\bibitem{dockercompose}
\BIBentryALTinterwordspacing
{Overview of Docker Compose}. Accessed 2019-08-21. [Online]. Available:
  \url{https://docs.docker.com/compose/}
\BIBentrySTDinterwordspacing

\bibitem{flask}
M.~Grinberg, \emph{Flask web development: developing web applications with
  python}.\hskip 1em plus 0.5em minus 0.4em\relax O'Reilly Media, 2018.

\bibitem{RestPerf}
J.~Fernandes, I.~Lopes, J.~Rodrigues, and S.~Ullah, ``Performance evaluation of
  restful web services and amqp protocol,'' in \emph{Fifth International
  Conference on Ubiquitous and Future Networks (ICUFN)}, 2018.

\bibitem{amqp}
{S. Vinoski}, ``Advanced message queuing protocol,'' \emph{IEEE Internet
  Computing}, no.~6, 2006.

\bibitem{rabbitmq}
\BIBentryALTinterwordspacing
{Messaging that just works - RabbitMQ}. Accessed 2019-08-21. [Online].
  Available: \url{https://www.rabbitmq.com}
\BIBentrySTDinterwordspacing

\bibitem{pony}
\BIBentryALTinterwordspacing
{PonyORM - Python ORM with beautiful query syntax}. Accessed 2019-08-21.
  [Online]. Available: \url{https://ponyorm.org}
\BIBentrySTDinterwordspacing

\bibitem{minio}
\BIBentryALTinterwordspacing
{MinIo} - object storage for {AI}. Accessed 2019-08-21. [Online]. Available:
  \url{https://min.io}
\BIBentrySTDinterwordspacing

\bibitem{archer}
\BIBentryALTinterwordspacing
{ARCHER}. Accessed 2019-08-21. [Online]. Available:
  \url{https://www.archer.ac.uk}
\BIBentrySTDinterwordspacing

\bibitem{cirrus}
\BIBentryALTinterwordspacing
{Cirrus}. Accessed 2019-08-21. [Online]. Available:
  \url{https://www.cirrus.ac.uk}
\BIBentrySTDinterwordspacing

\bibitem{beskow}
\BIBentryALTinterwordspacing
(2019) {Beskow}. [Online]. Available:
  \url{https://www.pdc.kth.se/hpc-services/computing-systems/beskow-1.737436}
\BIBentrySTDinterwordspacing

\bibitem{vestecnick}
N.~Brown, R.~Nash, G.~Gibb, B.~Prodan, M.~Kontak, V.~Olshevsky, and W.~{Der
  Chien}, ``\BIBforeignlanguage{English}{The role of interactive
  super-computing in using hpc for urgent decision making},'' in
  \emph{\BIBforeignlanguage{English}{High Performance Computing}}.\hskip 1em
  plus 0.5em minus 0.4em\relax Springer, 7 2019.

\end{thebibliography}

\end{document}